\documentclass{article}

\usepackage{amsmath}
\usepackage[final]{nips_2017}

\usepackage[utf8]{inputenc} 
\usepackage[T1]{fontenc}    
\usepackage{hyperref}       
\usepackage{url}            
\usepackage{booktabs}       
\usepackage{amsfonts}       
\usepackage{nicefrac}       
\usepackage{microtype}      
\usepackage{subfigure}
\usepackage{graphicx}
\usepackage{lineno}
\title{Atrial Fibrillation Detection and ECG Classification based on CNN-BiLSTM}

\author{
  Jiacheng Wang \\
  NYU\\
  \texttt{jw5728@nyu.edu} \\
   \And
  Weiheng Li\\
  NYU\\
  \texttt{wl1845@nyu.edu} \\
}

\begin{document}
\nolinenumbers
\maketitle
\begin{abstract}
It is challenging to visually detect heart disease from the electrocardiographic (ECG) signals. Implementing an automated ECG signal detection system can help diagnosis arrhythmia in order to improve the accuracy of diagnosis. In this paper, we proposed, implemented, and compared an automated system using two different frameworks of the combination of convolutional neural network (CNN) and long-short term memory (LSTM) for classifying normal sinus signals, atrial fibrillation, and other noisy signals. The dataset we used is from the MIT-BIT Arrhythmia Physionet. Our approach demonstrated that the cascade of two deep learning network has higher performance than the concatenation of them, achieving a weighted f1 score of 0.82. The experimental results have successfully validated that the cascade of CNN and LSTM can achieve satisfactory performance on discriminating ECG signals.

\end{abstract}

\section{Introduction}
The demand for efficiently monitoring subhealth problem increases rapidly. Cardiac disease is the leading cause of death nowadays. And one of the possibilities to detect, like what Apple Watch new version aim, is based on the real-time heartbeat remotely monitor cardiac health. Electrocardiograph is an easy and efficient way to define, detect and diagnose cardiac arrhythmia. And it is not hard for trained cardiologists to identify dozens of different types of heartbeats. However, researchers still could not successfully implement state-of-art supervised machine learning methods to achieve the same level of diagnosis. 

In medical, Cardiac Arrhythmia is characterized by heart rhythms that do not follow a normal sinus pattern. There are 12 leading types of abnormal arrhythmia rhythms
[1]. Atrial fibrillation is one of the most significant abnormal heart rhythm characterized by the rapid and irregular beating of our heart. The goal is to help mobile cardiac recording to classify Atrial Fibrillation from normal and other arrhythmia. Automated detecting cardiac arrhythmia can significantly help patients to get emergency treatment avoiding further troubles. For this type of task, accurately labeling different types of heartbeats from long term physiological activities is the key step for real-time diagnosis purposes. 

Traditional cardiac rhythms classification methods based on signal processing take advantage of frequency domain analysis to classify the difference among different types of cardiac. Also, some real-time modern filtering methods, e.g. wiener filter, were implemented to show time-domain analysis which includes instantaneous frequency, spectral entropy, and et, al. However, it takes time to hand-on extract these features from the original ECG signal and subject-wise difference makes it harder to standardize the classification procedures.
The novel solution of different types of cardiac classification from ECG signals implements machine learning algorithms, neural networks. Inspired from traditional work, researchers try to either extract features from 2-dimension spectrograph or extract from time-series 1-dimension signal. Based on the high dimensional feature vectors, the fully-connected layer and softmax layer can predict multi-class cardiac type. 

Previous researchers alluded deep convolutional neural network that has recently shown cardiologist-level performance for cardiac arrhythmia classification. These methods consume huge amounts of computation resources and use transfer knowledge frameworks from the traditional computer vision areas in object classification. In [2], it proposed to use ResNet to fully decomposite the real-time ECG signal and achieves cardiologist-level classification results. Also, the latest papers and researches use shallow CNN to analyze the spectrum graph generated based on signal processing study. energy/ entropy/ heartrate. Moreover, some paper[3][4] implements 2D CNN as the classifier relying on sets of hand-on features. The family of CNN models is proved to be suitable and capable of solving ECG classification tasks.

However, drawbacks from the CNN architectures are significantly obvious including fixed features and computation cost. Variation of ECG signals may not be represent properly by generic features. Researchers take their steps back to use neural networks for extracting temporal features. The Fourier transform of the spectrum inspires lots of researchers to implement recurrent neural network (RNN) to extract the long term and short term important features that mimic spectrum analysis. Both wider and deeper RNN were tested in previous work. Different types of RNN was compared in [5], and LSTM outperforms from GRU and RNN.

In this paper, we show how features extracted from pure ECG signals using the 1d-convolutional neural network (CNN) will perform much better than traditional hand-coded features which is much more efficient, transferable and accurate for representing the labeled signal and the classification tasks. Additionally, the novel recurrent neural network (RNN), especially our proposed bidirectional long-short term memory (BiLSTM) temporal feature extraction is powerful for using classifying and predicting time series data, which captures only important components compared with the latest frequency graph CNN based models. We developed our framework with the above two characteristics. Applied to data from the MIT-BIH arrhythmia database, our model outperforms state-of-art models with the highest f1 score among [6][7][8] models. Utilizing the characteristic of CNN and LSTM separately and combining these abilities together make our prediction reasonable and effective. 

The rest of this paper is organized as follows. In Section 2, the problem and algorithms are introduced. The methodologies including data preprocessing, model design and network architecture are described in detail in Section 3. Finally, Section 4 shows the performance of our model, concludes the paper and illustrates our future work.

\section{Problem Definition and Algorithms}
\subsection{Task Definition}
The primary goal of our project is to help classify cardiac arrhythmia which is characterized by heart rhythms that do not follow a normal sinus pattern. Due to the limitation of our dataset, we mainly focus on Atrial Fibrillation (aFib), one of the leading types of arrhythmia rhythm characterized by the rapid and irregular beating of our heart, classification among large sets of normal, Atrial Fibrillation, and other types of arrhythmia rhythms. The basic input of our project is one channel unfixed length electrocardiographic (ECG) signal and corresponding labels. And we preprocess and segment the basic input to fit in our proposed model, which will be specifically illustrated in section 4.1. After preprocessed, the original input data was segmented into 4 heartbeats with fixed intervals length, containing both before and after the R-peak timestamps, in case the final input of our model will be fixed-length heartbeats contained the same number of samples (1000 timestamps). The goal of our project, after training among our dataset, would be predicting aFib, normal and other types of arrhythmia labels for unlabeled ECG signals. Also, the validation part of prediction labels will be processed in the testing phase and compared to existing models and several different schemes from the previous work.
\subsection{Algorithms Description}
\subsubsection{Convolutional Neural Network}
Tradiontially, to solve the classification of cardiac arrhythmia (aFib in our case), many researchers focus on the study of sets of hand-crafted features including using RR intervals (time interval from one R wave to the next R wave), PR intervals, P-Wave absence and so on. However, because of the variants of ECG signals and the missing generic adaptation of underlying characteristics, a limited number of features from previous work may not be effective for signal detection. Therefore, we implied a convolutional neural network architecture, which was highly implemented in multiple different research domains, to extract spatial fusion features for discriminating aFib ECG signals out of normal ones. Feature maps were generated from the CNN output layer.

\subsubsection{Long-Short Term Memory}
There are various types of RNN in the research area, including vanilla RNN, shadow RNN, GRU, LSTM, and BiLSTM. Considering the characteristic of selectively retaining the history and current information in which highly will be useful information in detecting the difference among different ECG signal types, the bidirectional long-short term memory (BiLSTM) network was chosen as our basic architecture to extract features from a long time-series data with function defined as below:

\begin{equation}
\begin{aligned}
\overrightarrow{h_t} &=     \tanh{(W_{\overrightarrow{h_t}}x_t+W_{\overrightarrow{h_t}\overrightarrow{h_t}}h_{t-1}+b_{\overrightarrow{h_t}})}\\
\overleftarrow{h_t} &= \tanh{(W_{\overleftarrow{h_t}}x_t+W_{\overleftarrow{h_t}\overleftarrow{h_t}}h_{t-1}+b_{\overleftarrow{h_t}})}\\
y_{t} &= \tanh{(W_{\overrightarrow{h_0}\overrightarrow{h_t}}+W_{\overleftarrow{h_0}\overleftarrow{h_t}}+b_0)}
\end{aligned}
\end{equation}

Besides CNN, the BiLSTM is also capable of automatically extracting temporal features from ECG signals. The BiLSTM in our proposal posses the extracting ability for both ECG signals and generated CNN feature maps. 

\subsubsection{CNN-LSTM}
The Convolutional Neural Networks (CNNs) allow us to learn suitable feature representations from ECG signals rather than hand-on feature extraction. However, previous studies have shown the classification task based on the huge feature maps introducing the overfitting problem. Therefore, the Long-Short Term Memory (LSTM) units perform temporal dimensionality reduction for selecting the most useful features for the task of classification.

\subsubsection{Focal Loss}
It is obvious that our dataset is quite an imbalance for the ratio of normal recordings is more than $80\%$. Inspired by traditional computer vision study, the focal loss is implemented in our model. Focal loss is a modified version of cross-entropy (CE) with dynamically scales proposed by [9] where CE was defined below:
\begin{equation} 
    \text{CE}(\hat{y}) = -\log(\hat{y})
\end{equation}
The dynamic scaling factor is negative correlated with confidence. And focal loss (FL) was defined as below:
\begin{equation}
    \text{FL}(\hat{y}) = - (1-\hat{y})^{\gamma} \log(\hat{y}),\quad \gamma \geq 0
\end{equation}
where $\gamma$ is a focusing parameter. By implementing focal loss into our training model, we will reduce the contribution of normal samples and enhance the focus on samples which are difficult to classify. 
\section{Methodology}

\subsection{Data Preprocessing}

In clinical the multi-channel ECG signals were obtained. However, in our project, we obtain the annotation data from MIT-BIH Physionet ECG Dataset (2017 competition version). In which, holds 8258 one channel individual recordings with two blindly annotated labels from two professional cardiologists including normal, atrial fibrillation, and other (simplified with N, aFib, and O). 

\begin{figure}[ht]
    \centering
    \subfigure[After Standardization ECG Signals]{
        \includegraphics[scale=0.8]{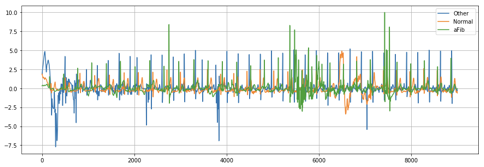}
        \label{fig1.1}}
    \subfigure[After Denoising ECG Signals]{
        \includegraphics[scale=0.8]{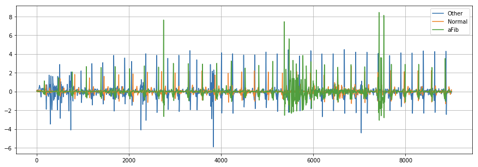}
        \label{fig1.2}}
    \caption{ECG Signals with fixed length (9000 timestamps)}
\end{figure}

Each recording varies from 9000 timestamps to 18000 timestamps with the same sample frequency 300Hz and the amplitudes for each recording range from -10 mV to 10 mV. We standardized all the recordings with zero mean shown in Fig 1.a. Observed with noisy signals, a bandpass filter will be first implied to denoise the original signal avoiding unexpected error or bias shown in Fig 1.b. The bandpass frequency was chosen based on previous work and set it as 3Hz and 45Hz for the range of normal human signals. In order to standardize the input of the model and to reduce the training time, sequences longer than 9000 timestamps were truncated down.

From previous clinical studies, Q, R, and S (QRS complex) are three deflections to represent a single heartbeat even in a period ECG signal. The time of its occurrence and its shape provide valuable information about the state heart. Traditional methods use wavelet transformation, frequency analysis, and digital filter, which extract local maximum value, to detecting R peaks. And R peaks indices are proved as critical classification indicators both for human and computer-aided. In our method, we would not include indicators other than R peaks indices (we calculate the distance between two adjacency R peaks and call it R-R intervals) to use our model to extract features from pure ECG signals. A novel dual-slope QRS detection algorithm proposed by [xx], consist of the fact that the fastest change of the slope happens at the peaks and provide low computational complexity and high accuracy. Therefore, we implement this algorithm in our model for detecting R peaks (shown in Figure 2.a)and calculate the R-R intervals for obtaining the attention features in the classification tasks.

\begin{figure}[ht]
    \centering
    \subfigure[R-peak Detection (Normal)]{
        \includegraphics[scale=0.45]{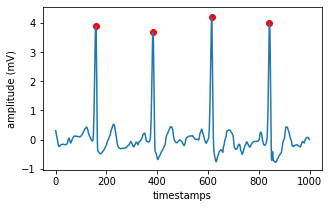}
        \label{fig2.1}}
    \subfigure[Normal Segment]{
        \includegraphics[scale=0.45]{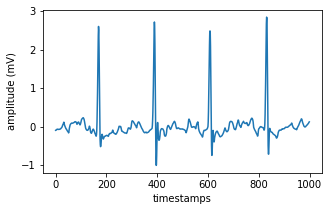}
        \label{fig2.2}}
    \subfigure[aFib Segment]{
        \includegraphics[scale=0.45]{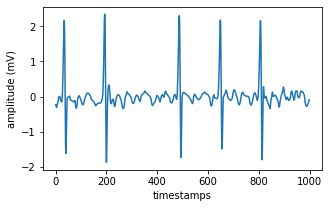}
        \label{fig2.3}}
    \subfigure[Other Segment]{
        \includegraphics[scale=0.45]{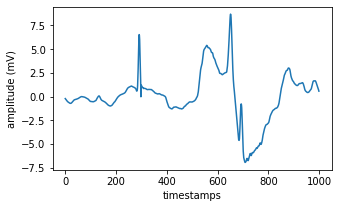}
        \label{fig2.4}}
    \caption{ECG Segments}
\end{figure}

After computing R peaks, we derived an algorithm to split the ECG signals of 9000 lengths into various segments containing at least 4 R-R intervals (the AFib signals usually contain more than 4 R peaks) with a fixed length of 1000. The typical segments of normal, atrial fibrillation and other signals are shown in Figure 2.b, 2.c, and 2.d.

It is very common to use sets of features for signal processing problems instead of using original signal itself. However, according to [MIT active learning applied to patient-adaptive heartbeat classification], it mentioned that the inter- and intra-patient differences will hugely influence the accuracy of ECG classification result using features-based methods. Therefore, in our feature extraction part, we only apply the R-R intervals into original ECG signals to get segmentation and followed model will deal with feature extraction part. The overall proposed data preprossing part was summarized in block diagram Figure 3. which return sets of fixed length (1000 Timestamps) ECG segments and will be the input our model illustrated in the next section.
\begin{figure}[ht]
    \centering
    \includegraphics[scale = 0.7]{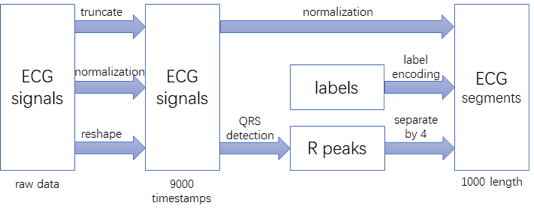}
    \caption{Data Frame work}
    \label{fig3}
\end{figure}

\subsection{System Architecture}

\begin{figure}[ht]
    \centering
    \subfigure[CNN Concat LSTM]{
        \includegraphics[scale=0.6]{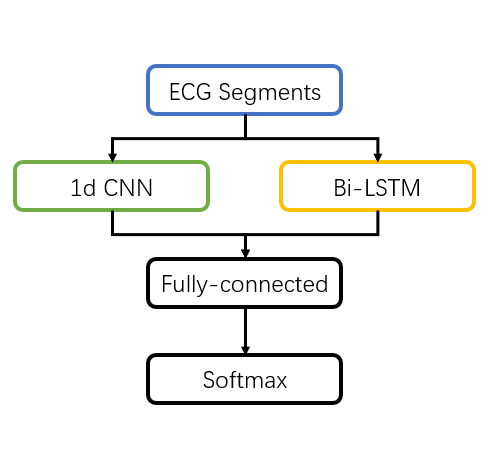}
        \label{fig4.1}}
    \subfigure[CNN Feed LSTM]{
        \includegraphics[scale=0.6]{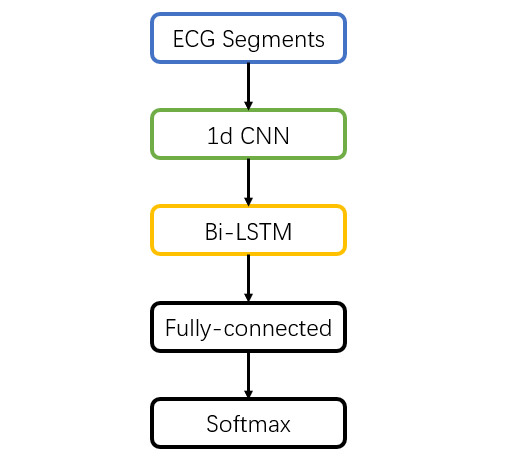}
        \label{fig4.2}}
    \caption{System Architecture}
\end{figure}

In this work, two different CNN-LSTM networks are compared and implemented to detect the atrial fibrillation signals among normal signals and other noisy signals automatically with ECG segments. These proposed networks are hybrid deep neural networks that combine CNN with LSTM either horizontally or vertically. Figure 4 shows the architecture of the proposed model and gives an overview of the framework in detail. Regardless of the different layouts, the composition of both CNN portions and LSTM portions are identical in these two models. Figure 5 shows the hyperparameters of the convolutional neural network.

Layers 1 to 4 of the CNN are convolutional layers coupled with normalization layers and dropout layers, whereas the first two layers are also coupled with max-pooling layers. An ECG signal of 1000 length is transformed into a feature vector of 256 lengths after passing this network. The LSTM portion is bidirectional, coupled with dropout and its hidden state is set to be 100 such that an ECG signal of 1000 length is transformed into a feature vector of 100 lengths.

\begin{figure}[ht]
    \centering
    \includegraphics[scale=0.5]{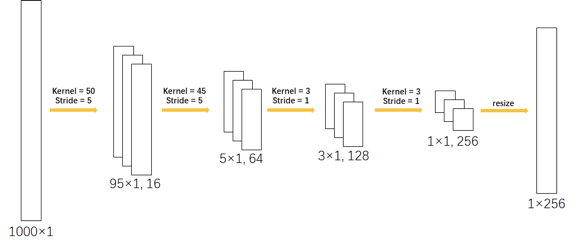}
    \caption{Hyperparameters of CNN}
    \label{fig5}
\end{figure}

\subsubsection{CNN Concat LSTM}
In this network, either CNN or LSTM is regarded as a feature extractor that is implemented to capture the spatial information and temporal feature respectively. It concatenates the spatial feature and temporal feature generated by different deep learning models directly. A fully-connected layer follows in order to predict the output. The output of the fully-connected layer then is passed into a softmax function and is transformed as the probability of belonging corresponding classes.

\subsubsection{CNN Feed LSTM}
In this model, the LSTM receives the output of CNN and extracts features in a deeper way. The output of the CNN portion, which is the spatial feature vector, is fed to LSTM to extract the temporal information. It is not a simple combination of feature vectors but a cascade of deep learning networks. The output of the LSTM portion is passed into a fully-connected layer to predict the class. The softmax function is applied to transform the output into the form of probability.

\subsection{Metrics Definition}
In order to get rid of the influence of data imbalance, the performance of the model is evaluated based on the f1 score that combines precision and sensitivity. Additionally, the specificity is chosen to be the second criteria of the model since the topic is of diagnosis. A model with good performance should retain a high f1 score and a high specificity. The formula of f1 score and specificity is displayed below:
\begin{equation}
    SP=\frac{TN}{TN+FP}
\end{equation}
\begin{equation}
    F1=\frac{2TP}{2TP+FN+FP}
\end{equation}

\subsection{Train and Validation}
To prevent the model from overfitting and tracing the training process, $70\%$ of the training data is actually used to train the model, while $30\%$ of the training data is used to validate the performance of the network at the end of each epoch. The model is trained for 50 epochs using the back-propagation algorithm with a batch size of 2048. The learning rate is set to 0.001 and is also applied to Adam optimizer so as to accelerate the learning process. The total training and validation process was processing under the GPU acceleration environment.

\section{Conclusion and Future Work}
\subsection{Model Performance}
In total, three different schemes are tested. The first scheme is a trained BiLSTM and the other two are CNN Concat LSTM (Scheme A) and CNN Feed LSTM (Scheme B) which are mentioned above. The validation accuracy curves of our three schemes are presented in Fig 6.b. It is obvious to observe that both scheme A and scheme B have better performance than a single BiLSTM network. Especially for Scheme B, it achieves the highest f1 score of 0.82 after training 50 epochs comparing with the other two models. Moreover, our experiment results show that all our proposed LSTM based models perform more effectively arrhythmia detection results compared with other previous hand-on features based, and CNN-only models, all of which provide only up to average 70\% f1 scores.

The confusion matrices for Scheme A and Scheme B are presented in Figure 7.a and Figure 7.b. It is apparent to conclude that Scheme B performs better than Scheme A on classifying normal signals and aFib signals that satisfies our goal. The specificity of Scheme A and Scheme B are calculated to be 0.966 and 0.978 respectively based on confusion matrices. 

\begin{figure}[ht]
    \centering
    \subfigure[Loss]{
        \includegraphics[scale=0.55]{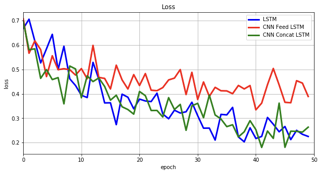} \label{fig6.1}}
    \subfigure[Validation Scores]{
        \includegraphics[scale=0.55]{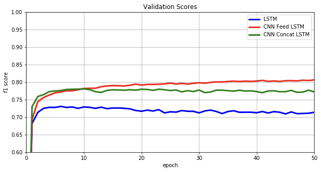}
        \label{fig6.2}}
    \caption{Loss and validation scores of models}
    
\end{figure}

\begin{figure}[ht]
    \centering
    \subfigure[CNN Feed LSTM]{
        \includegraphics[scale=0.5]{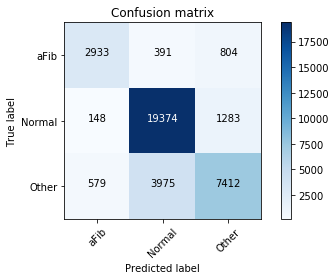}
        \label{fig7.1}}
    \subfigure[CNN Concat LSTM]{
        \includegraphics[scale=0.5]{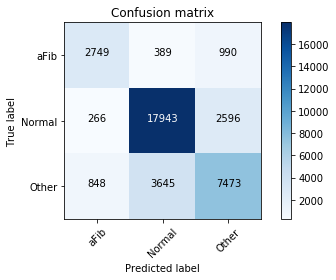}
        \label{fig7.2}}
    \caption{Confusion matrix}
    
\end{figure}

\subsection{Conclusion}
An effective and efficient atrial fibrillation detection model is proposed in this paper. The 256-dimensional feature vector from 4 heartbeats (4 R-R intervals of ECG signal) is extracted for each fixed-length ECG segments.  Also, the f1 score and specificity show the effectiveness of our model. Notably for the cardiac type ‘Other’ where we noticed that it is more often to be misclassified from the other two types. The network did not perform well to generate the Other class as this label consists of multiple different cardiac arrhythmias where the variability inter-class problem is arising. Comparably, the Normal class is easily discriminated offering the best scores from the other two classes because it retains the represented shape among every sample.

\subsection{Future Work}
In future work, we are considering time-series synthesis heard from the guest lecture. There are several successful works on soundtrack and music synthesis. However, the part in ECG signal synthesis is still under research. One of the latest work[10] has tried to use CNN and LSTM to be the discriminator for the traditional audio generator. However, the performance of the result still could not capture the variation of different types of patients. 
 
We also noticed that, sequence-to-sequence LSTM generator is powerful in NLP domain[11]. Some researchers added Attention layer to implement auto-Encoder/Decoder. Several experiments of auto-encoder/decoder were tested during our project. However, it is hard to generate the common starting vector for individual subject.

\section*{Acknowledgments}
    Thanks to Prof.Crstina Savin to help us get the credential to access MIT-BIH ECG Dataset[12] and all her advice and suggestions on our project. Appreciate DS-3001.1 course TAs for providing excellent time series analysis python examples for helping us get used to this topic. Thanks to Prof.Joan Bello's guest lecture for proving music synthesis inspiring us to do future work on ECG synthesis learning.
    
\section*{Contribution}
The code is available on \url{https://github.com/liweiheng818/ECG-Signal-Analysis}

Weiheng focuses on network setup and debug, while Jiacheng focuses on data preprocessing. And we work on the previous paper review and our project paper write-up together.

\section*{References}
[1] Haraldsson, H.\ \& Edenbrandt, L.\ (2004) Detecting acute myocardial infarction in the 12-lead ECG using Hermite expansions and neural networks. \  {\it Artificial Intelligence in Medicine}, pp.\ 127-136. 

[2] Rajpurkar, P.\ \& Ng, A. Y.\ (2017) Cardiologist-level arrhythmia detection with convolutional neural networks. {\it arXiv preprint} arXiv:1707.01836.

[3] Jun, T. J.\ \& Kim, Y. H.\ (2018). ECG arrhythmia classification using a 2-D convolutional neural network. {\it arXiv preprint} arXiv:1804.06812.

[4] Lu, W.\ \& Chu, J.\ (2018). Feature fusion for imbalanced ECG data analysis. {\it Biomedical Signal Processing and Control} 41, 152-160.

[5] Salloum, R., Schnell, E.\ \& Kuo, C. C. J.\ (2017) ECG-based biometrics using recurrent neural networks. {\it In 2017 IEEE International Conference on Acoustics, Speech and Signal Processing (ICASSP)} pp. 2062-2066.

[6] Clifford, G. D.\ \& Mark, R. G. \ (2017, September). AF Classification from a short single lead ECG recording: the PhysioNet/Computing in Cardiology Challenge 2017. {\it In 2017 Computing in Cardiology (CinC)} pp. 1-4.

[7] Xiong, Z.\ \& Zhao, J.\ (2017, September). Robust ECG signal classification for detection of atrial fibrillation using a novel neural network. {\it In 2017 Computing in Cardiology (CinC)} pp. 1-4.

[8] Zihlmann, M.\ \& Tschannen, M.\ (2017, September). Convolutional recurrent neural networks for electrocardiogram classification. {\it In 2017 Computing in Cardiology (CinC)} pp. 1-4.

[9] Ballotta, E.\ \& B9racchini, C.\ (2010). Predictors of electroencephalographic changes needing shunting during carotid endarterectomy. {\it Annals of vascular surgery} 24(8), 1045-1052.

[10] Zhu, F.\ \& Shen, B.\ (2019). Electrocardiogram generation with a bidirectional LSTM-CNN generative adversarial network. {\it Scientific reports} 9(1), 6734.

[11] Malhotra, P.,\ \& Shroff, G.\ (2016). LSTM-based encoder-decoder for multi-sensor anomaly detection. {\it arXiv preprint} arXiv:1607.00148.

[12] Moody, G. B.\ \& Mark, R. G. \ (2001). The impact of the MIT-BIH arrhythmia database. {\it IEEE Engineering in Medicine and Biology Magazine} 20(3), 45-50.
\end{document}